\documentclass[fleqn]{mnras}

\usepackage{newtxtext,newtxmath}
\usepackage[T1]{fontenc}

\usepackage{graphicx}	% Including figure files
\pdfoutput=1

\title[NGC~5846\_UDG1]{
Stellar Velocity Dispersion and Dynamical Mass of the Ultra-Diffuse
Galaxy NGC 5846\_UDG1 from the Keck Cosmic Web Imager}

\author[D. A. Forbes et al.]{Duncan A. Forbes,$^{1}$\thanks{E-mail: dforbes@swin.edu.au}
Jonah S. Gannon$^{1}$, Aaron J. Romanowsky$^{2,3}$, Adebusola Alabi$^{3}$
\newauthor
Jean P. Brodie$^{1,3}$, 
Warrick J. Couch$^{1}$ and Anna Ferr\'e-Mateu$^{1,4}$
\\
$^{1}$Centre for Astrophysics \& Supercomputing, Swinburne University, Hawthorn VIC 3122, Australia\\
$^{2}$Department of Physics \& Astronomy, San Jos\'e State University, San Jose, CA 95192, USA\\
$^{3}$University of California Observatories, 1156 High St., Santa Cruz, CA 95064, USA\\
$^{4}$Institut de Ciencies del Cosmos (ICCUB), Universitat de Barcelona (IEEC-UB), E02028 Barcelona, Spain
}

\date{Accepted XXX. Received YYY; in original form ZZZ}

\pubyear{2020}

\begin{document}
\label{firstpage}
\pagerange{\pageref{firstpage}--\pageref{lastpage}}
\maketitle

\begin{abstract}
    The ultra-diffuse galaxy in the NGC~5846 group (NGC~5846\_UDG1) was shown to have a large number of globular cluster (GC) candidates from deep imaging as part of the VEGAS survey. Recently, Muller et al. published a velocity dispersion, based on a dozen of its GCs.
    %, of  $\sigma_{GC}$ = 9.4$^{+7.0}_{-5.4}$ km/s. 
    Within their quoted uncertainties, the resulting dynamical mass allowed for either a dark matter free or a dark matter dominated galaxy. Here we present spectra from KCWI which reconfirms membership of the NGC~5846 group and reveals a stellar velocity dispersion for UDG1 of $\sigma_{GC}$ = 17 $\pm$ 2 km/s. Our dynamical mass, with a reduced uncertainty, indicates a very high contribution of dark matter within the effective radius. We also derive an enclosed mass from the locations and motions of the GCs using the Tracer Mass Estimator, finding a similar mass inferred from our stellar velocity dispersion.  
    We find no evidence that the galaxy is rotating and is thus likely pressure-supported. The number of confirmed GCs, and the total number inferred for the system ($\sim$45), suggest a total halo mass of $\sim2 \times 10^{11}$ M$_{\odot}$. A cored mass profile is favoured when compared to our dynamical mass. Given its stellar mass of 1.1$\times$10$^{8}$ M$_{\odot}$, NGC~5846\_UDG1 appears to be an ultra-diffuse galaxy with a dwarf-like stellar mass and an overly massive halo.  
    
\end{abstract}


\section{Introduction}

Ultra-diffuse Galaxies (UDGs) were first identified using a novel telescope (actually a collection of multiple cameras called The Dragonfly Array) optimised for detecting low surface brightness objects by van Dokkum et al. (2015). Further deep imaging has revealed UDGs in all environments, although they are more common in clusters (e.g. Janssens et al. 2019). They are defined to have half-light radii (R$_e$) $>$ 1.5 kpc and central surface brightnesses ($\mu_0$) $>$ 24 mag. per sq. arcsec in the g band. With stellar masses (M$_{\ast}$) of around 10$^8$ M$_{\odot}$, they have half-light radii comparable to the Milky Way but the stellar content of dwarfs. However, perhaps their most remarkable property is that some appear to have overmassive (Beasley et al. 2016; 
van Dokkum et al. 2016; Forbes et al. 2019) and even undermassive dark matter halos (Danieli et al. 2019; van Dokkum et al. 2019a). Total halo masses are either inferred from counting their globular clusters (GCs) and applying the scaling relation for normal galaxies (Burkert \& Forbes 2019) or they are inferred from measured dynamical masses and scaled assuming a mass profile (Wasserman et al. 2019; Gannon et al. 2020). 

Although the best studied UDG with an overmassive halo, Dragonfly~44 (DF44) in the Coma cluster, obeys the GC number vs halo mass relation, the undermassive halo UDG DF2 in the NGC~1052 group does not (although in this case the compact sources seen may not all be GCs). 
The alternative approach is to measure the  velocity dispersion of a UDG and calculate its  dynamical mass. 
Due to observational challenges, this has only been carried out for a small number of UDGs by either measuring the integrated properties of the galaxy stars or from the collective motions of their GCs. The number of UDGs with {\it both} stellar and GC orbit-based velocity dispersions in the literature is currently only two (i.e. NGC~1052--DF2 and VCC~1287). Clearly more UDGs need to have their stellar velocity dispersions measured in order to test whether the GC derived values, often based on very small samples, are reliable.  

Integral field units (IFUs) optimised for spatially-resolved spectroscopy of low surface brightness objects are now available on 8-10m class telescopes. They include KCWI on Keck, MUSE on VLT and Megara on GTC. Each has its own comparative advantage for studying UDGs: KCWI has an excellent blue response and can operate at high spectral resolution; MUSE has a large field-of-view ($\sim$1 $\times$ 1 sq. arcmin) which allows for on-chip sky subtraction of more distant UDGs, Megara offers high spectral resolution and dedicated fibres for sky subtraction. Even with long exposure times on such large telescopes, these IFUs are often used in `light-bucket' mode where all spaxels are collapsed to give a single kinematic measure for the galaxy.

NGC~5846\_UDG1 was first identified by Mahdavi et al. (2005) as a "very low surface brightness galaxy" which they dubbed NGC~5846-156. From deeper VEGAS imaging (Spavone et al. 2017), Forbes et al. (2019) determined it had the properties of a UDG if at the distance of the NGC~5846 group. The imaging also revealed a system of at least 20 compact objects (likely to be GCs). More recently, Muller et al. (2020; hereafter M20), using MUSE on the VLT, measured a recession velocity placing it in the NGC~5846 group. They also confirmed the association of a dozen GCs with the UDG (which they refer to as MATLAS J15052031+0148447). From the motions of these objects they measured a velocity dispersion 
%of $\sigma_{GC}$ = 9.4$^{+7.0}_{-5.4}$ km/s 
and determined an enclosed dynamical mass within the de-projected half-light radius (following Wolf et al. 2010) of 1.8$^{+3.7}_{-1.5}$ $\times$ 10$^{8}$ M$_{\odot}$. Using their luminosity, this gave a mass-to-light ratio in the V band of 4.2$^{+8.6}_{-3.4}$. Their lower limit is 
comparable to the galaxy's stellar M/L and draws comparison to the two UDGs in the NGC~1052 group that appear to lack dark matter (van Dokkum et al. 2019), whereas their higher limit suggests a substantial dark matter halo as seen in some other UDGs (e.g. Gannon et al. 2020). 
%similar to DF44 (van Dokkum et al. 2016). 
As their GC sample is subject to small number statistics (see Laporte et al. 2019)
%and uncertain inclination effects. 
%, the globular clusters 
%revealed a possible rotation signature which affects the dynamical mass calculation as an assumption of no rotation was assumed. 
a high precision stellar velocity dispersion is desirable in order to better define the mass-to-light ratio of 
NGC~5846\_UDG1 and hence test whether the populous GC system seen in the imaging of Forbes et al. (2019) is associated with a massive halo as might be expected.

Here we present spectra obtained for NGC~5846\_UDG1 using KCWI on the Keck telescope from which we measure its stellar velocity dispersion.  
KCWI has the advantage of superior spectral resolution over MUSE, i.e. $\sigma_{inst}$ = 13 km/s for KCWI vs 45 km/s for MUSE. We also discuss other UDGs for which both stellar and GC-orbit based velocity dispersions are available. We search for signs of rotation in the main body of the galaxy and measure the recession velocity for two bright GCs. We derive a dynamical mass and compare it to an inferred total halo mass for NGC~5846\_UDG1. 

\section{Observations and Data Reduction}

\begin{figure}
    \centering
    \includegraphics[width = 0.49 \textwidth]{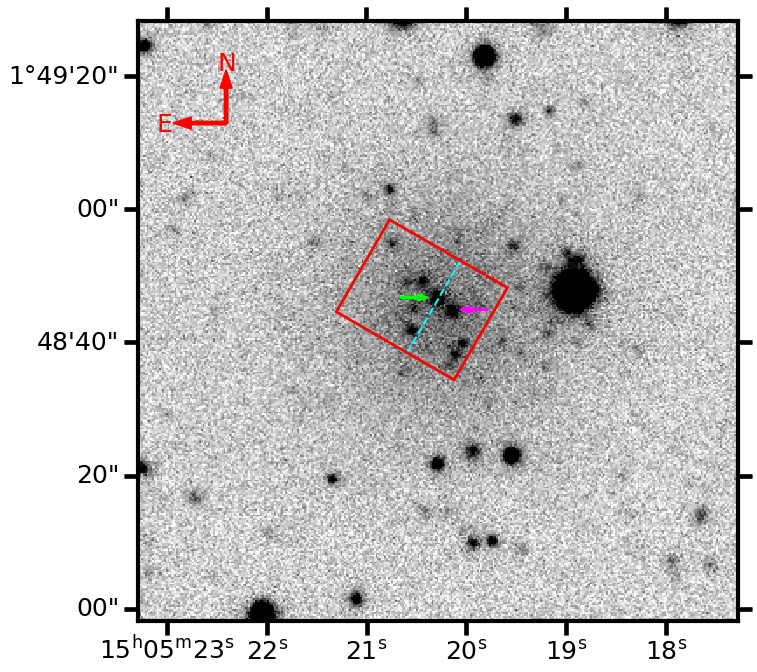}
    \caption{Image of NGC~5846\_UDG1 with overlay showing KCWI footprint. The solid red overlay shows the orientation of KCWI's 16 $\times$ 20 sq. arcsec field-of-view at PA = 60$^{o}$. The dashed cyan line shows the axis used to separate the NE from SW regions of the galaxy. The green arrow points to globular cluster GC9 and the purple arrow arrow to GC10. We extract spectra of the galaxy and the two GCs separately. The galaxy has a half-light radius of 17.7 $\pm$ 0.5 arcsec with its centre near GC9 (which may be the stellar nucleus). }
    \label{fig:fov}
\end{figure}

NGC~5846\_UDG1 was observed with KCWI (Martin et al. 2010) on the Keck II telescope under program N061 on the nights of 2019 March 30, May 1 and May 29. We employed the BH3 grating centred at 5110~\AA~ with a full wavelength range of 4850 to 5350~\AA~.
The spectral resolution is  $\sigma_{inst}$ = 13 km/s. We used the medium slicer with a field-of-view of 16 $\times$ 20 sq. arcsec, with the long axis of KCWI orientated roughly in the NE-SW direction with a PA of 60 degrees (see Fig.~\ref{fig:fov}). 
Our field-of-view extends to around 50\% of the half-light radius of 17.7 $\pm$ 0.5 arcsec (Forbes et al. 2019). 
Conditions were dark but some thin cloud was present. The total time on the science target was 21900 sec. We alternated on-target observations with offset positions, obtaining a total of 19200 sec on sky. A standard star was observed with the same setup and used for flux calibration. The data were reduced following the method of Gannon et al. (2020), i.e. we ran the standard KCWI data but skipped stage 5 (standard sky subtraction). 
Following this we perform the data correction described in appendix A of Gannon et al. (2020) and sky subtraction using our PCA-based routine. After application of the corresponding barycentric corrections, we median combine the sky-subtracted spectra. 

\begin{table}
%	\centering
	\caption{NGC~5846\_UDG1 Properties}
	\label{tbl:UDG1}
	\footnotesize
	\begin{tabular}{lc}
		\hline
Property & Value\\
\hline
RA (J2000)    & 15:05:20\\                
Dec (J2000)      & 01:48:47\\
Velocity (km/s) & 2167 $\pm$ 2 \\
Distance (Mpc)      & 24.89\\            
R$_e$ (kpc)     & 2.14 $\pm$ 0.06\\
Sersic n & 0.68 \\
%M$_g$ (mag) & -14.2 $\pm$ 0.2\\
M$_V$ (mag) & -14.5 $\pm$ 0.2\\
g--i (mag) & $\sim$1\\
%$\mu_e$ (mag/arcsec$^2$) & 26.0 $\pm$ 0.05 \\
$\mu_0$ (mag/arcsec$^2$) & 24.8 $\pm$ 0.1 \\
M$_{\ast}$ (M$_{\odot}$) & 1.1 $\times$ 10$^8$\\
$\sigma_{\ast}$ (km/s) & 17 $\pm$ 2\\
M$_{1/2}$ (10$^8$ M$_{\odot}$) & 5.46 $\pm$ 1.3 \\
M$_{1/2}$/L$_{1/2}$ & 19.9 $\pm$ 5.7 \\
GCs & $\sim$45\\
\hline
	\end{tabular}
	
	Notes: values from this work or Forbes et al. (2019).
	
\end{table}

\begin{figure}
    \centering
    \includegraphics[width = 0.49 \textwidth]{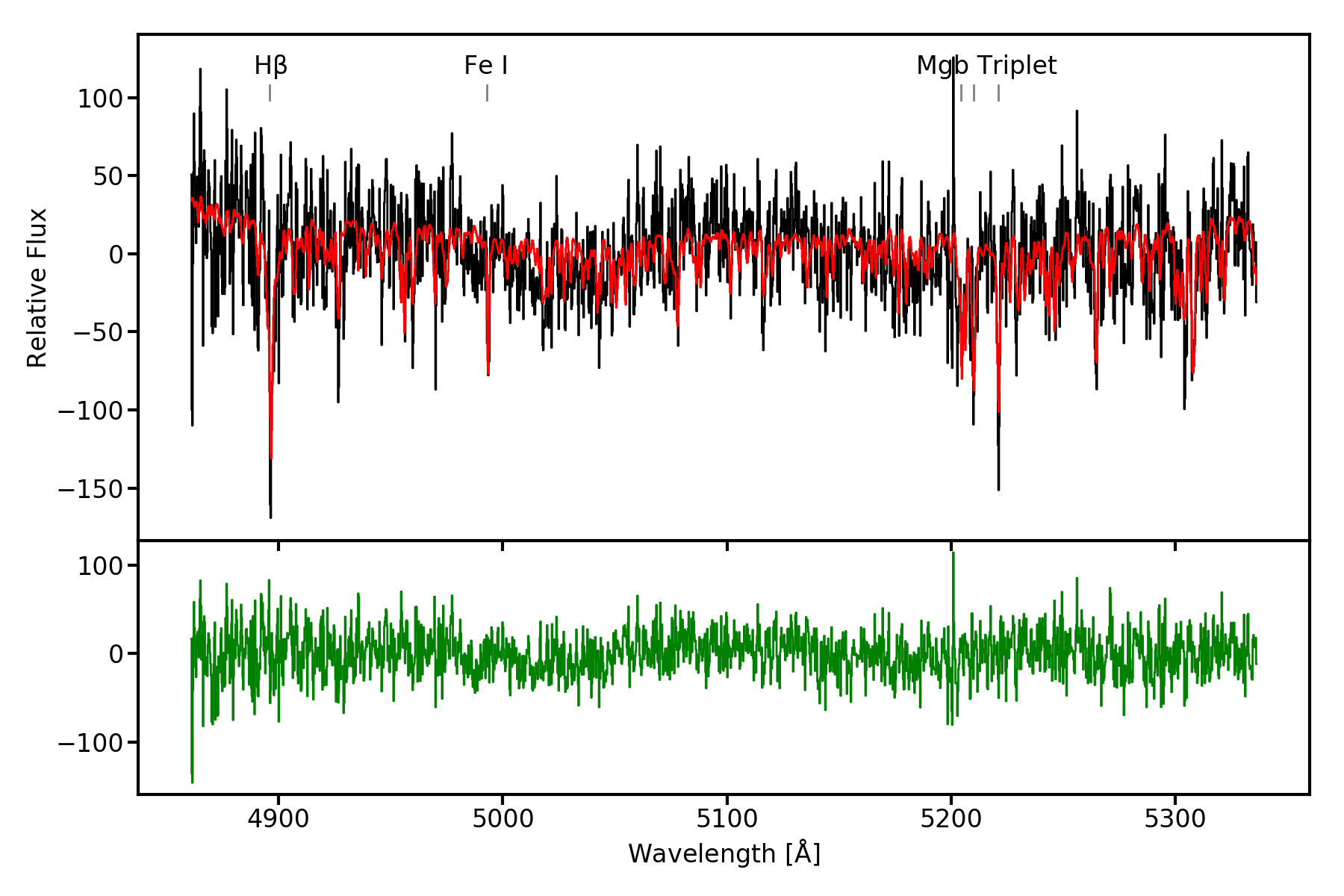}
    \caption{Our final spectrum for NGC~5846\_UDG1 with a representative stellar population model fit. Our final spectrum (black) is over-plotted with a representative fit (red) from our exhaustive \texttt{pPXF} fitting with the Coelho (2014) synthetic stellar library. Residuals of the fit (green) are shown in the lower panel. H$\mathrm{\beta}$, Fe I and the Mg b triplet have been labelled. Based on our fitting of the NGC~5846\_UDG1 spectrum we recover a velocity dispersion of $\sigma_{\ast}$ = 17 $\pm$ 2 km/s.}
    \label{fig:spectrum}
\end{figure}

After masking the two brightest compact sources present (the remaining compact sources in the field-of-view contribute $\le$6\% of the total flux) we extract a spectrum for the galaxy.  
In Fig.~\ref{fig:spectrum} we show the final spectrum obtained by collapsing all spectra across the galaxy. It has 
a S/N of $\sim$20 per \AA\ in the continuum.
We also show the best fit to the spectrum using \texttt{pPXF} (Cappellari 2017) and the resulting residuals. In order to fully explore the possible choices of \texttt{pPXF} input parameters we fit a wide ranging selection of input parameters as described in Gannon et al. (2020). We do so using both the Coelho (2014) synthetic stellar library and an observation of the Milky Way GC M3. We observed M3 with the same KCWI instrumental configuration, although with a slightly lower central wavelength to allow for the predicted disparity in recessional velocities between M3 and the target. We also fit 5 different spectral regions of the final spectrum to ensure our final value is not driven by any particular spectral region. Fitting with both the Coelho (2014) and M3 templates displays good convergence in both recessional velocity and velocity dispersion. Our final values for recessional velocity and velocity dispersion are 2167 $\pm$ 2 km/s and $\sigma_{\ast}$ = 17 $\pm$ 2 km/s respectively. 
%These values are taken from the average of all fits with the Coelho (2014) library. 

\section{Results}

The measured properties of NGC~5846\_UDG1 from this work and the deep imaging study of Forbes et al. (2019) are summarised in Table 1. 

Our measurement of the recession velocity (2167 $\pm$ 2 km/s) is similar to that quoted by M20 in their table 1, i.e. 2156 $\pm$ 9.4 km/s. This velocity places NGC~5846\_UDG1 in the NGC 5846 group, which has a distance of 25 Mpc based on a mean of the surface brightness fluctuation and fundamental plane distances (Tully et al. 2013). Here we assume the same distance as used by Forbes et al. (2019) and Spavone et al. (2017) of 24.89 Mpc (M20 assumed a distance of 26.3 Mpc.) 

We also extract spectra for the two bright compact sources in the KCWI field-of-view. Using the naming convention from Forbes et al. (2019), they are GC10 and GC9. 
%The latter is the brightest compact source and may be the stellar nucleus of NGC~5846\_UDG1 (Forbes et al. 2019). 
We measure a recession  velocity for GC10 of 2163 $\pm$ 3 km/s compared to 2147 $\pm$ 8 km/s by M20 (which they call GC5), and for GC9 we measure 2166 $\pm$ 3 km/s compared to 2147 $\pm$ 5 km/s in M20 (which they call GC6). We note that 
%the velocity differences are within one KCWI pixel and that 
GC9 (the brightest and centrally located source) 
has a similar velocity to the host galaxy (2167 $\pm$ 2  km/s) and may be its stellar nucleus (see also Forbes et al. 2019). 

\subsection{Kinematics}

We measure a stellar velocity dispersion within our KCWI field-of-view 
%(8.4 arcsec $\times$ 20.4 arcsec) 
of $\sigma_{\ast}$ = 17 $\pm$ 2 km/s. Our uncertainty is the error on the mean value from fits using different stellar libraries, templates and spectral regions as described in Section 2.
Within the joint uncertainties the GC based velocity dispersion of M20, $\sigma_{GC}$ = 
9.4$^{+7.0}_{-5.4}$ km/s is consistent with our value. 
M20 measured old ages and low metallicities typical of MW GCs, although we question their quoted levels of uncertainty for some individual GCs. For example, they quoted an age of 9.1 Gyr with an accuracy of +1.6 and -2.6 Gyr for a 
a GC spectrum with a S/N per pixel of only 3.7. 
We note that the work of Conroy et al. (2018) suggests that a S/N
$>$40\AA~ is needed to measure age and metallicity to
within 0.1 dex
Unfortunately, they did not provide spectra nor their model fit in their paper. 
We also note that if they had used the galaxy velocity as a prior in their MCMC analysis  of the GC system, then they may have derived a slightly higher velocity dispersion (see their fig. 6). 
%We also note that renamed the 20 GC candidates listed by Forbes et al. (2019), i.e. GC1...GC20. For example, the brightest GC of M20 (GC6) is actually GC9 from Forbes et al. (2020). 

\begin{figure}
	\centering
	\includegraphics[width=0.80\linewidth, angle=-90]{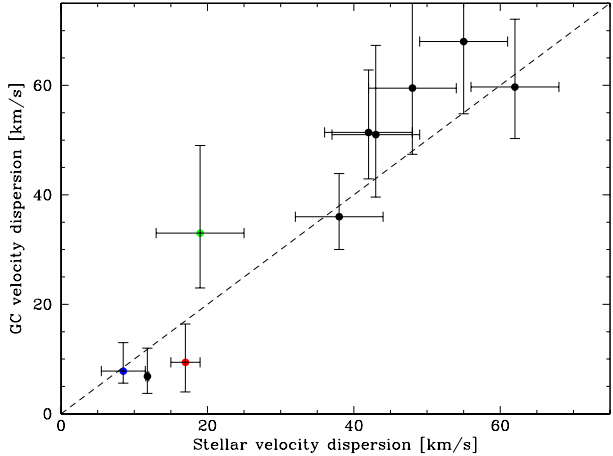}
	\caption{Comparison of velocity dispersions from stars and globular cluster systems. The dashed line is not a fit but a unity line. Ultra-diffuse galaxies are shown by coloured symbols (red for  NGC5846\_UDG1, green for VCC~1287 and blue for DF2). Other early-type galaxies (from Laporte et al. 2019 and Fahrion et al. 2020) are shown in black. 
	}
	\label{fig:sigma}
\end{figure}

In Fig.~\ref{fig:sigma} we compare velocity dispersion measures from stars with those from GCs for NGC~5846\_UDG1 and for other galaxies in the literature. There are only two other UDGs for which both stellar and GC-based velocity dispersions exist, i.e. VCC~1287 (Gannon et al. 2020) and NGC~1052--DF2 (Danieli et al. 2019). For these galaxies, and NGC~5846\_UDG1, we summarise their values from the two methods in Table 2. 
For DF2 both low velocity dispersions are very consistent with each other. For NGC~5846\_UDG1 the stellar velocity dispersion is {\it higher} than the GC one, whereas for VCC~1287 it is {\it lower} than the GC one, 
{\it although in both cases the values are consistent within the quoted joint uncertainties. } The stellar velocity dispersions listed in Table 2  have lower quoted uncertainties than those from GCs. 
%We note that if the velocity outlier in the VCC~1287 GC system is removed a re-calculated velocity dispersion is much lower at $\sigma_{GC}$ = 21 km/s. 
We also include the Fornax dSph galaxy in Fig.~\ref{fig:sigma} with the GC velocity dispersion of 4 GCs from Laporte et al. (2019) and the stellar velocity dispersion from McConnachie et al. (2012). In addition, we include some dwarf early-type galaxies from the Fornax~3D survey (Fahrion et al. 2020). We do not expect a perfect match between the stellar and GC-based velocity dispersions given differences in spatial coverage, anisotropies, tracer distributions etc. However, 
within the quoted uncertainties, velocity dispersions measured from GC systems and integrated stars for these low mass galaxies are generally consistent with each other. One notable exception is VCC~1287 which has a much reduced GC velocity dispersion of $\sigma_{GC}$ = 21 km/s if the velocity outlier were removed (see Gannon et al. 2020). 
We note that for large early-type galaxies with rich GC systems, 
the two velocity dispersion measures are well correlated on a one-to-one basis (Pota et al. 2013). 
%To remove the influence of small number statistics 
%on future GC studies, one should of course attempt to obtain as many radial %velocities in a given GC system as possible. 

\begin{table}
%	\centering
	\caption{UDG Velocity Dispersions}
	\label{tbl:UDG1}
	\footnotesize
	\begin{tabular}{lccc}
		\hline
Name & $\sigma_{\ast}$ & $\sigma_{GC}$ & Ref.\\
     & (km/s) & (km/s) & \\
\hline
NGC~5846\_UDG1 & 17 $\pm$ 2 & 9.4$^{+7.0}_{-5.4}$ & 1,2\\
VCC~1287 & 19 $\pm$ 6 & 33$^{+16}_{-10}$ & 3,4\\
NGC~1052-DF2 & 8.5$^{+2.3}_{-3.1}$ & 7.8$^{+5.2}_{-2.2}$ & 5,6\\
\hline
	\end{tabular}
	
Notes: 1 This work, 2 Muller et al. 2020, 3 Gannon et al. (2020), 4 Beasley et al. (2016), 5 Danieli et al. (2019), 6 van Dokkum et al. (2018).

\end{table}

\subsection{Masses}

The absolute magnitude of NGC~5846\_UDG1 from Forbes et al. (2019) is M$_V$ = --14.5. Adjusting to our adopted distance of 24.89 Mpc, M20 measured a magnitude M$_V$ = --14.88, whereas Mahdavi et al. (2005) measured --14.0 (assuming a typical UDG colour V--R = 0.3; Forbes et al. 2020). Thus our magnitude lies between these two published values. Our total luminosity, L$_V$ = 0.55 $\times$ 10$^{8}$ M$_{\odot}$,  translates to a total stellar mass of M$_{\ast}$ = 1.1 $\times$ 10$^8$ M$_{\odot}$ for a 
stellar M/L$_V$ ratio = 2 (M20 have confirmed that NGC~5846\_UDG1 is old and metal-poor). 

We follow Wolf et al. (2010) to calculate the 
enclosed dynamical mass. The Wolf et al. formula minimises the effect of non-isotropic orbits on the derived mass, while assuming a pressure-supported system (see Section 3.3 below). We assume that 
our velocity dispersion measurement is representative of the mean value within the de-projected half-light radius of r$_{1/2}$ = 4/3 $\times$ R$_e$ = 2.85 kpc. If the velocity dispersion profile is actually rising with radius, as seen in DF44 (van Dokkum et al. 2019b), then the true enclosed mass will be greater. The Wolf, or half, mass is M$_{1/2} = 930 \sigma^2 R_e \sqrt{b/a}$, where the ellipticity $b/a$ = 0.9 from M20. 
We derive the mass within the deprojected half-light radius to be M$_{1/2}$ = 5.46 ($\pm$ 1.3)
%5.75 ($\pm$ 1.35) 
$\times$ 10$^8$ M$_{\odot}$. From the luminosity above, the resulting mass-to-light ratio is 
M$_{1/2}$/L$_{1/2}$ = 19.9 $\pm$ 5.7. 
%20.9 $\pm$ 4.9 

M20 applied the Wolf et al. formula to their GC-based velocity dispersion (assuming the GC system had the same effective size as the galaxy half-light radius). They quoted a half mass of 
M$_{1/2}$ = 1.8$^{+3.7}_{-1.5}$ $\times$ 10$^8$ M$_{\odot}$. The main reason for their lower half mass is the lower velocity dispersion of their GCs compared to the integrated stars of the gaalxy.  They also 
derived a mass-to-light ratio M$_{1/2}$/L$_{1/2}$ = 4.2$^{+8.6}_{-3.4}$. Using their mass and our luminosity above, this ratio would rise to 6.5.
%$^{+13.3}_{-5.3}$. 

\begin{figure*}
    \centering
    \includegraphics[width = 0.98 \textwidth]{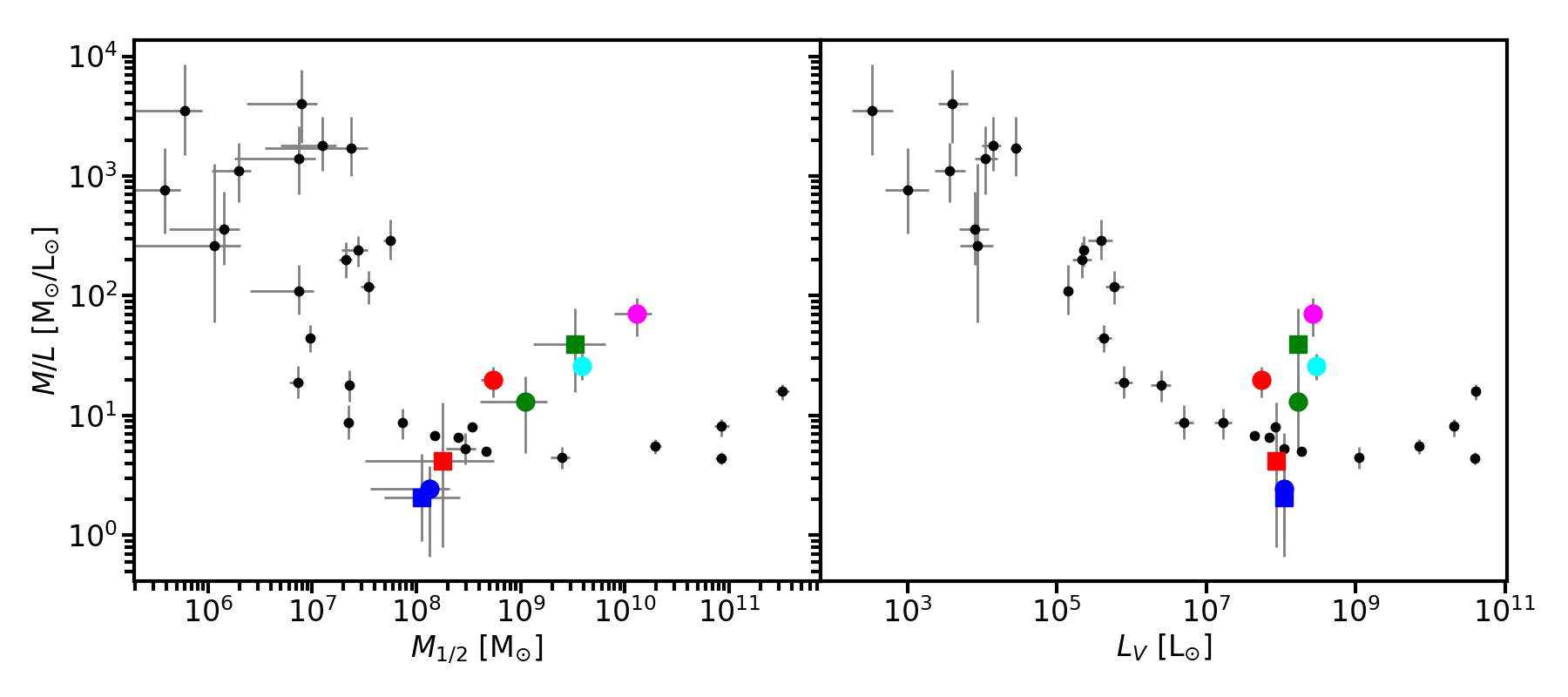}
    \caption{Dynamical mass-to-light ratio (M$_{1/2}$/L$_{1/2}$) vs dynamical mass within the deprojected half-light radius ({\it left}) and total V band luminosity ({\it right}). Ultra-diffuse galaxies are shown by coloured symbols (red for  NGC5846\_UDG1, green for VCC~1287 and blue for DF2, cyan for DF44 and magenta for DGSAT~I). For NGC5846\_UDG1, VCC~1287 and DF2 the plots show the mass-to-light ratio derived from GC (squares) and stellar velocity dispersions (circles). 
    For NGC~5846\_UDG1 we show mass and luminosity taken directly from M20 (red square) and this work (red circle).
    The black symbols show Local Group dwarfs and other pressure-supported galaxies from  
    Wolf et al. (2010) and Forbes et al. (2011). Both distributions have a local minimum around 10$^9$ in solar units, with most UDGs found above the distribution of normal galaxies with higher mass-to-light ratios. 
    }
    \label{fig:ML}
\end{figure*}

In Fig.~\ref{fig:ML} we show the half mass (M$_{1/2}$) to half light (L$_{1/2}$) ratios for UDGs and other objects from the literature.  For three UDGs we show M$_{1/2}$ estimates based on both stellar velocity dispersions and GC-based ones. For other galaxies, including two well-studied UDGs
(DGSAT~I; Mart{\'\i}n-Navarro et al.(2019)
and DF44; van Dokkum et al. 2019b), only mass-to-light ratios from stellar velocity dispersions are available.
The plot shows the well-known U-shape for normal galaxies (from Wolf et al. 2010 and Forbes et al. 2011) with UDGs mostly scattering above the trend to higher mass-to-light ratios. The exceptions are NGC~1052-DF2 (for which both GC and stellar derived masses indicate little or no dark matter) and the GC based mass for NGC~5846\_UDG1 from M20. For a discussion of aperture effects on the mass-to-light we refer the reader to van Dokkum et al. (2019).

An alternative approach for calculating the enclosed mass is to use the Tracer Mass Estimator (TME) of Watkins et al. (2010) based on the positions and velocities of the GCs. The TME has been used successfully by Alabi et al. (2016) to measure the enclosed mass within 5~R$_e$ for the early-type galaxies in the SLUGGS survey using samples of tens to hundreds of GCs. 
Based on the virial theorem, the TME requires assumptions regarding the tracer distribution, gravitational potential and orbits quantified by $\gamma$, $\alpha$ and $\beta$ respectively. van Dokkum et al. (2017) examined the radial distribution of GCs in two GC-rich UDGs (DF44 and DFX1). They found that the distribution was more extended than the galaxy light and could be fitted by a Sersic profile with n $\sim$ 3. 
%$^{+0.6}_{-0.9}$. 
The profile (their fig. 2) can be approximated by a  power-law slope of $\sim$1 within R$_e$ and hence a deprojected 3D slope of around 2. Thus we adopt $\gamma$ = 2. 
%We note that the Milky Way's GC system has a 3D slope of 3.5. 
For the potential we explore the range --1 $<$ $\alpha$ $<$ 1 noting that an isothermal potential ($\alpha$ = 0) may be appropriate for low surface brightness galaxies (de Blok et al. 2002).
We allow orbits to range from mildly radial ($\beta$ = +0.5) to tangential ($\beta$ = --0.5), with isotropic orbits having $\beta$ = 0. Applying the TME to the GCs in M20, we derive a dynamical mass enclosed within the outermost GC (i.e. 2.2 kpc) of 3.48 ($\pm$ 0.84)  
%^{+1.2}_{-1.7}$ 
$\times$ 10$^8$ M$_{\odot}$, with the 
uncertainty given by the range in potential and orbital anisotropy described above. 
Our TME dynamical mass is larger than the GC-based half mass, but agrees within the uncertainties. 
Allowing for a reasonable range in $\gamma$ would increase the uncertainty to around $\pm$50\%; where assuming a  
lower (higher)  value of $\gamma$ would lead to a reduced (increased) TME mass. 
The resulting mass-to-light ratio from our TME analysis is M$_{1/2}$/L$_{1/2}$ = 12.7 $\pm$ 3.7. 
%19.3$^{+4.4}_{-6.4}$. 

Forbes et al. (2019) detected 20 GC candidates associated with NGC~5846\_UDG1.
%, a dozen of which were subsequently confirmed by M20 to be bona fide members. 
If we assume the same mean GC luminosity function turnover magnitude (i.e. M$_V$ = --7.3) and width as found by Miller \& Lotz (2007) for dE galaxies in the Virgo cluster, then the inferred total GC population is around 45. 
%4 $\pm$ 7 (the uncertainty is based on estimates of the statistical contamination rate). 
For this count, the Burkert \& Forbes (2019) relation between the number of GCs and the halo mass implies a total halo mass of $\sim$2 $\times$ 10$^{11}$ M$_{\odot}$. 
Given that there are several unknowns in translating the detection of GC candidates into an estimate of the total GC system, it is useful to note that M20 confirmed at least a dozen GCs associated with NGC~5846\_UDG1. This gives a lower limit on the total halo mass with this relation, of 0.6 $\times$ 10$^{11}$ M$_{\odot}$. Both of these total halo mass estimates far exceed the total stellar mass of 
1.1 $\times$ 10$^8$ M$_{\odot}$. For comparison, M33 has the same inferred total halo mass of 
2 $\times$ 10$^{11}$ M$_{\odot}$ (Seigar et al. 2011) but its stellar mass is some 50$\times$ larger. Thus 
NGC~5846\_UDG1 can be described as having 
a dwarf-like stellar mass with an overly massive dark matter halo.

\begin{figure}
	\centering
	\includegraphics[width=0.99\linewidth, angle=0]{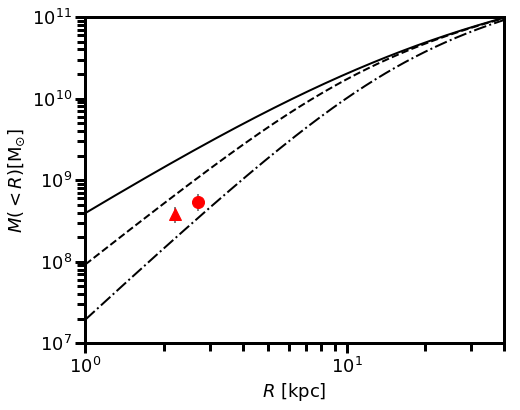}
	\caption{Halo mass profiles. Three halo mass profiles (Navarro et al. 1996; solid, Di Cintio et al. 2014; dashed and Burkert 1995; dot-dashed) corresponding to a total halo mass of  $\sim$2 $\times$ 10$^{11}$ M$_{\odot}$ are shown. The half mass (derived from integrated stars) and the TME mass (derived from globular cluster orbits) are shown as a red circle and red triangle respectively. The Di Cintio et al. (2014) and Burkert (1995) mass profiles with cores are more consistent than a cuspy NFW profile with our dynamical mass measurements.
		}
	\label{fig:halo}
\end{figure}

Assuming a total halo mass of 2 $\times$ 10$^{11}$ M$_{\odot}$ as described above, we can compare our half mass and TME mass with different halo mass profiles. We chose three profiles, namely the universal NFW profile (Navarro, Frenk \& White 1996), the mass profile resulting from the dwarf galaxy simulations of Di Cintio  et al. (2014), which resembles a core-NFW profile, and the observationally-motivated mass profile for dwarf galaxies from Burkert (1995). For details of the NFW and Di Cintio et al. profiles see the appendix of Di Cintio et al. (2014). 
We show these three profiles, with the same total halo mass, compared to our half and TME masses measured at small radii in Fig.~\ref{fig:halo}. The plot shows that our mass measurements lie between the Di Cintio and Burkert mass profiles for a total halo mass as given by the total number of GCs. This suggests that NGC~5846\_UDG1 may be better represented by a core mass profile as predicted in some simulations (e.g. 
Carleton et al. 2019) than a cuspy one. However, we cannot, at this stage, rule out other mass profiles and/or different total halo masses. Similar conclusions, favouring a cored mass profile, were reached by Gannon et al. (2020) for VCC~1287, and by Wasserman et al. (2019) for DF44. 

\subsection{Does NGC~5846\_UDG1 Rotate?}

An assumption of both the half mass and the TME mass calculations is that the system is dominated by random motions, and not bulk rotation. The GC system of NGC~5846\_UDG1 shows some evidence for rotation (see fig. 9 in M20). However, after fitting a $\it sine$ 
function and solving for a variable rotation axis and rotation amplitude, via an MCMC analysis, M20 could not confirm (nor rule out) bulk rotation of the GC system. 
A rotational contribution to the velocity would scale as $(V_{rot}/sin~i)^2$ and so any correction to the half mass or TME mass would be highly dependent on the unknown inclination $i$. 
We note that the galaxy itself appears rather circular on the sky.
%so defining the true rotation axis is problematic.
%We note that the main body of NGC~5846\_UDG1 is almost circular on the sky.

%, and implies that both the half mass and TME mass, under the no rotation assumption, are both upper limits. M20 estimate that the half mass could be lower by as much as 2/3 if the GC system were seen edge-on. 

In order to search for rotation in the stellar body of NGC~5846\_UDG1 we have split our KCWI dataset into two parts,  with data coming from the NE and SW regions of the galaxy (see Fig.~\ref{fig:fov} illustrating the two regions). Each part has a S/N $\sim$ 16 per \AA~ and again we mask out the two bright GCs. We effectively partition the galaxy across a position angle of 150 degrees. Our position angle lies within the range of the rotation axis of the GC system as determined by M20, i.e. 110 $\pm$ 50 degrees. We note that visual inspection of a simple 2D velocity map of the GCs relative to the host galaxy systemic velocity does not reveal an obvious rotation axis.

From the NE region of the galaxy we measure a recession velocity of 2166 $\pm$ 2 km/s and from the SW region 2171 $\pm$ 2 km/s. Thus we find little, or no, evidence for rotation along this axis. 
We also measure similar velocity dispersions in the two regions, both of which are consistent with the overall value we measure (see Table 1). We note that the best studied UDG, with a radial kinematic profile, is that of DF44 in the Coma cluster. In this case rotation of the stellar body can be strongly ruled out (van Dokkum et al. 2019). A general lack of rotation in the UDG population would argue against formation mechanisms that predict rotation in UDGs (Amorisco \& Loeb 2016; Wright et al. 2020; 
Cardona-Barrero et al. 2020).

\section{Conclusions}

In this work, using spectra from KCWI, we reconfirm membership of the NGC~5846 group and measure the stellar velocity dispersion for the ultra-diffuse galaxy NGC~5846\_UDG1. Our stellar velocity dispersion of $\sigma_{\ast}$ = 17 $\pm$ 2 km/s is consistent within the uncertainty of the previously published value based on a dozen globular clusters (i.e. $\sigma_{GC}$ = 9.4$^{+7.0}_{-5.4}$ km/s). However our value, with a lower uncertainty, translates into a mass-to-light ratio that indicates a significant fraction of dark matter 
within the deprojected half-light radius. 
Applying the Tracer Mass Estimator to the GC system also suggests the presence of dark matter at small radii.

We compare velocity dispersions derived from GC systems and integrated stars for pressure-supported galaxies from the literature, finding that the current data are generally consistent with a one-to-one relation. However, given the relatively small number of GCs associated with a given UDG and 
various systematic uncertainties present (e.g. orbits, inclination, rotation), we favour the use of integrated stars to probe UDG kinematics and their dynamical masses.

By dividing our KCWI data into two regions along  the NE-SW direction we test for bulk rotation in the galaxy, finding none. The galaxy appears to be dominated by random motions. This argues against formation mechanisms that predict clear rotation in UDGs. 
We also measure the recession velocity for the two brightest compact objects in our field-of-view (one of which may be the galaxy stellar nucleus given its brightness and central location). We find them both to have a similar velocity to that of the host galaxy. 

Our previous deep imaging, and spectroscopy in the literature, revealed a populous globular cluster system associated with NGC~5846\_UDG1. We estimate a total GC system count of $\sim$45. The scaling relation between the number of GCs and halo mass implies a total halo mass of $\sim$2 $\times$ 10$^{11}$ M$_{\odot}$. 
This suggests NGC~5846\_UDG1 hosts an overly massive dark matter halo for its stellar mass (1.1 $\times$ 10$^{8}$ M$_{\odot}$. 
Given this total halo mass, Di Cintio et al. (2014) and Burkert (1995) mass profiles with cores are more consistent than a cuspy NFW profile with our dynamical mass measurements.
%Comparison with the galaxy's stellar mass of $\sim$10$^{8}$ M$_{\odot}$ suggests 
%This suggests NGC~5846\_UDG1 hosts an overly massive dark matter halo for its stellar mass. %Thus NGC~5846\_UDG1 shares similarities to DF44 in the Coma cluster, i.e. it has a dwarf-like stellar mass but a giant-like dark matter halo. 
Further studies are required to determine the overall frequency of UDGs with overly massive halos, whether they are non-rotating and if they lie within cored dark matter halos.  

\section*{Acknowledgements}

We thank E. Iodice for helpful comments and S. Danieli for help with the observations. 
DAF thanks the ARC for financial support via DP160101608. 
AFM has received financial support through the Postdoctoral Junior Leader Fellowship Programme from La Caixa Banking Foundation (LCF/BQ/LI18/11630007). 
AJR was supported by National Science Foundation grant AST-1616710 and as a Research Corporation for Science Advancement Cottrell Scholar.

This work was partially supported by a NASA Keck PI Data Award, administered by the NASA Exoplanet Science Institute.
The data presented herein were obtained at the W. M. Keck Observatory, which is operated as a scientific partnership among the California Institute of Technology, the University of California and the National Aeronautics and Space Administration. The Observatory was made possible by the generous financial support of the W. M. Keck Foundation. The authors wish to recognize and acknowledge the very significant cultural role and reverence that the summit of Maunakea has always had within the indigenous Hawaiian community.  We are most fortunate to have the opportunity to conduct observations from this mountain. 

\section{Data Availability}

The data are available via the Keck Observatory Archive (KOA): https://www2.keck.hawaii.edu/koa/public/koa.php 18 months after observations are taken.

\end{document}